\documentclass[journal]{IEEEtran}

\usepackage{cite}
\usepackage[pdftex]{graphicx}
\graphicspath{{./figs/}}
\DeclareGraphicsExtensions{.pdf,.png,.jpg}
\usepackage{amsmath}
\usepackage{mathrsfs}
\usepackage{pifont}
\usepackage{algorithmic}
\usepackage{array}
\usepackage{multirow}
\ifCLASSOPTIONcompsoc
  \usepackage[caption=false,font=normalsize,labelfont=sf,textfont=sf]{subfig}
\else
  \usepackage[caption=false,font=footnotesize]{subfig}
\fi
\usepackage{stfloats}
\ifCLASSOPTIONcaptionsoff
 \usepackage[nomarkers]{endfloat}
 \let\MYoriglatexcaption\caption
 \renewcommand{\caption}[2][\relax]{\MYoriglatexcaption[#2]{#2}}
\fi
\usepackage{url}
\usepackage{pgfplots}
\pgfplotsset{compat=1.13}
\usepackage{diagbox}
\usepackage{threeparttable}
\usepackage{makecell}

\begin{document}

\title{Progressive Joint Low-light Enhancement and Noise Removal for Raw Images}

\author{Yucheng Lu and Seung-Won Jung, \IEEEmembership{Senior Member,~IEEE}

\thanks{\textit{Corresponding author: Seung-Won Jung.}}
\thanks{Y. Lu is with the Department of Multimedia Engineering, Dongguk University, Seoul, Korea. S.-W. Jung is with the Department of Electrical Engineering, Korea University, Seoul, Korea.}
\thanks{This research was supported in part by the Basic Science Research Program through the National Research Foundation of Korea (NRF) funded by the Ministry of Science, ICT Future Planning (NRF-2020R1F1A1069009), and in part by the National Research Foundation of Korea (NRF) grant funded by the Korea government (MSIT) (No. 2020R1A4A4079705).}
}

\markboth{IEEE TRANSACTIONS ON IMAGE PROCESSING}%
{Shell \MakeLowercase{\textit{et al.}}: Bare Demo of IEEEtran.cls for IEEE Journals}

\maketitle

\begin{abstract}
Low-light imaging on mobile devices is typically challenging due to insufficient incident light coming through the relatively small aperture, resulting in low image quality. Most of the previous works on low-light imaging focus either only on a single task such as illumination adjustment, color enhancement, or noise removal; or on a joint illumination adjustment and denoising task that heavily relies on short-long exposure image pairs from specific camera models. These approaches are less practical and generalizable in real-world settings where camera-specific joint enhancement and restoration is required. In this paper, we propose a low-light imaging framework that performs joint illumination adjustment, color enhancement, and denoising to tackle this problem. Considering the difficulty in model-specific data collection and the ultra-high definition of the captured images, we design two branches: a coefficient estimation branch and a joint operation branch. The coefficient estimation branch works in a low-resolution space and predicts the coefficients for enhancement via bilateral learning, whereas the joint operation branch works in a full-resolution space and progressively performs joint enhancement and denoising. In contrast to existing methods, our framework does not need to recollect massive data when adapted to another camera model, which significantly reduces the efforts required to fine-tune our approach for practical usage. Through extensive experiments, we demonstrate its great potential in real-world low-light imaging applications.
\end{abstract}


%
\IEEEpeerreviewmaketitle

\section{Introduction}
\IEEEPARstart{S}{martphone} cameras dominate in terms of the global market share for all digital cameras. On top of them being highly compact and easy-to-use, the convenience of direct photo sharing through the internet means most people nowadays capture daily moments via phone cameras. In most cases where the surroundings are sufficiently bright, such as in sunny outdoor scenery, the images produced by these phone cameras have satisfactory quality. Unfortunately, this cannot be guaranteed when it comes to low-light imaging: First, phone cameras are usually equipped with a relatively small aperture, which limits the number of photons received by the sensor, resulting in low intensity for most pixels; Second, during the conversion from analog signals to digitized pixel values in image signal processing (ISP), various types of noise are introduced; Third, to avoid motion blur introduced by handshaking, the exposure time should be within the safe shutter speed \cite{yuan2007image}. Consequently, the images obtained in low-light environments often suffer from heavily underexposed regions, evident noise, and dull color, which not only fail to deliver sufficient information to the viewers but also affect the performance of high-level computer vision tasks such as object detection \cite{li2021photon}.

Some efforts have been made to improve the quality of low-light images on hardware and software. While the former tries to improve the sensor sensitivity or add artificial light sources such as infrared and ultraviolet flashes \cite{Krishnan09}, the latter does not need extra costs and thus is more popular. Some publicly available photo editors, such as Lightroom, provide a rich set of settings for users to manually alter photos according to their preferences. Although the results can be of higher aesthetic quality, the editing process requires solid backgrounds that are not friendly for non-professionals. The significant time and efforts spent on manual adjustments also make it impractical for online social interactions. In addition to manual tools, there are several built-in photo enhancers in smartphones that can process images automatically. However, these are designed for minor alterations and thus are less helpful for low-light enhancement where more tremendous efforts are required.

Compared with general enhancements for well-lit images, quality improvement for low-light images is more challenging. Non-learning-based studies \cite{yuan2012automatic, wang2013naturalness, wang2016fusion} that mainly focused on handcrafted priors yielded limited performance. A recent trend is thus applying convolutional neural networks (CNNs) to learn a better non-linear mapping, which has shown a tremendous performance boost in enhancing global and local contents. Nevertheless, there are still certain areas that can be improved: Most of the previous methods are designed for only a single task, such as illumination enhancement or color enhancement, while the aesthetic quality is more than just a single aspect. Also, denoising is either missing, which is not aligned with the fact that sensor-specific noise universally appears in low-light images and may vary a lot between cameras, or heavily relies on paired training data for its removal, which is limited to the specific camera used for making the dataset and not directly applicable when being adapted to other cameras.

We claim that an excellent low-light image should inherit the following characteristics: (a) sufficient illumination; (b) adequate contrast; (c) enhanced color; and (d) low noise level. In this paper, we present a learning-based framework that performs joint enhancement and denoising for low-light images to meet these requirements. We design a two-branch structure that learns enhancement and denoising features separately but simultaneously applies them to the input image. Remarkably, the coefficient estimation branch operates in a noise-insensitive low-resolution space and predicts parameters that can be used to enhance input images with arbitrary resolution. The joint operation branch operates in a full-resolution space and performs camera-specific denoising along with enhancement progressively. To make full use of several existing datasets, we design a strategy that trains the coefficient estimation branch in a weakly supervised manner and the joint operation branch iteratively. By this approach, the proposed framework successfully learns a wide range of highly non-linear operations, resulting in high-quality low-light images. Moreover, the two-branch design takes advantage of low computational complexity and easy applicability to other camera models without the need to recollect a massive amount of training data. In summary, our contributions are as follows:

\begin{itemize}
    \item We present a framework that performs joint enhancement and denoising of low-light images. The proposed framework results in images with higher quality compared to current state-of-the-art methods both qualitatively and quantitatively.
    \item We design a two-branch structure that estimates enhancement parameters in low-resolution via bilateral learning and applies joint enhancement and denoising in full-resolution progressively. This design enables the former to have a large receptive field insensitive to noise while preserving high-resolution features for the latter.
    \item We propose a strategy that uses several existing datasets developed for enhancement and denoising, respectively. For the coefficient estimation branch, a carefully designed cost function combined with zero-reference loss and high-level perceptual color loss enables weakly supervised network learning. For the joint operation branch, a more accurate noise model estimated using only a few dark image samples is employed in data synthesis.
\end{itemize}

The rest of this paper is organized as follows: Section II reviews previous research related to our work; Section III describes the proposed framework and the details of each component; Section IV analyzes the proposed framework and presents the experiment results in comparison with state-of-the-art methods; Section V concludes the paper and discusses our future work.

\section{Related Works}
This section presents previous works related to the problems we address in this paper. We grouped them into three categories: image enhancement, image denoising, and joint enhancement and denoising.

\subsection{Image Enhancement}
Images taken in low-light environments often contain poor visibility and dull color. The former is the consequence of insufficient illumination, and the latter depends on the illumination distribution of the captured scene. These images can be enhanced by extracting the illumination and reflectance components according to the Retinex theory \cite{land1977retinex}. Many methods have been proposed to recover these two components: Guo \textit{et al.} \cite{guo2016lime} proposed an optimization framework to extract the illumination map and enhance it using gamma correction. Zhang \textit{et al.} \cite{zhang2018high} designed a series of handcrafted constraints and used relative total variation to extract a structure-insensitive illumination map. Zhang \textit{et al.} \cite{zhang2019dual} generated a normal illumination map along with its inverted version and then fused them to obtain the final result. 

Besides the Retinex theory-based methods, it is feasible for networks to learn a direct mapping with the help of paired data. Inspired by the bilateral guided upsampling \cite{chen2016bilateral}, Gharbi \textit{et al.} \cite{gharbi2017deep} proposed a lightweight network that learns a series of enhancement operations. Huang \textit{et al.} \cite{huang2019hybrid} proposed a Laplacian enhancing unit to enhance image features under different scales and combined the pixel-wise scaling and addition operations into final image reconstruction. Similarly, Afifi \textit{et al.} \cite{afifi2020learning} built a coarse-to-fine exposure correction method using overexposed and underexposed image pairs. Ren \textit{et al.} \cite{ren2019low} proposed a hybrid network that handles global contents and local details separately. Moran \textit{et al.} \cite{moran2020deeplpf} applied a set of differentiable filters to the initial enhancement output from a U-Net-like network. Wang \textit{et al.} \cite{wang2020lightening} proposed a network consisting of several lightning back-projection blocks that iteratively perform the enhancement.

The training of the above methods all requires lots of paired images, which are time-consuming and laborious to obtain. Hence, learning from unpaired data has also attracted considerable attention. To achieve aesthetic-driven color enhancement, Deng \textit{et al.} \cite{deng2018aesthetic} employed a generative adversarial network (GAN) \cite{goodfellow2014generative} to supervise the aesthetic quality of generated images. Chen \textit{et al.} \cite{chen2018deep} proposed a two-way GAN with ideas borrowed from CycleGAN \cite{zhu2017unpaired}. Jiang \textit{et al.} \cite{jiang2019enlightengan} employed a relativistic discriminator \cite{jolicoeur2018relativistic} in their dual-discriminator design, yielding more natural images with enhanced details.

\subsection{Image Denoising}
Low-light images heavily suffer from a low signal-to-noise ratio (SNR) due to insufficient photons. To overcome this limitation, some works paid particular attention to single-photon sensors \cite{gupta2019asynchronous}: Ingle \textit{et al.} \cite{ingle2019high} proposed a passive free-running method to capture temporal recordings at random and non-uniformly spaced time intervals, achieving extremely-high dynamic range. Chi \textit{et al.} \cite{chi2020dynamic} utilized a deblurring teacher and a denoising teacher to train the student network for dynamic scene restoration, while Ma \textit{et al.} \cite{ma2020quanta} proposed a method that wraps the binary frames using the estimated coarse-to-fine patch flows and reconstructs the result via robust frequency-domain merging. Elgendy \textit{et al.} \cite{elgendy2021low} introduced a framework that demodulates color through frequency selection and transfers luma features to help reconstruct chroma channels. As to noise reduction for conventional sensors, burst imaging offers an effective way to improve SNR. Although some efforts \cite{hasinoff2016burst, godard2018deep, liba2019handheld} have been made, the time spent on taking and merging images still set a burden to more practical applications. 

Consequently, single image denoising has gained great attention. While early works based on image priors \cite{buades2005non,elad2006image} showed limited performance for low-light images, recent studies on dense CNNs have greatly improved the denoising quality and reached state-of-the-art, as reported in \cite{abdelhamed2019ntire, abdelhamed2020ntire}. With the sources of noise have been well-studied \cite{konnik2014high}, there are some works dedicated to noise modeling: Foi \textit{et al.} \cite{foi2008practical} proposed a joint Poisson and Gaussian model that describes the signal-dependent noise as well as the signal-independent noise. Hasinoff \textit{et al.} \cite{hasinoff2010noise} combined the two forms of noise in one single heteroscedastic Gaussian model. Several denoisers \cite{brooks2019unprocessing, marras2020reconstructing, guan2019node} trained using these noise models have reported state-of-the-art performance. Besides, Hirakawa \textit{et al.} \cite{hirakawa2011skellam} addressed the Poisson noise estimation in Haar wavelet and Haar filterbank transform domains and proposed their Skellam shrinkage method. Zhang \textit{et al.} \cite{zhang2017improved} proposed a Poisson mixture model to capture the long-tail nature of sensor noise. Similarly, Wei \textit{et al.} \cite{wei2020physics} employed the Tukey lambda distribution to obtain a more precise noise model, while Abdelhamed \textit{et al.} \cite{abdelhamed2019noise} combined the basic noise parameters with a neural network, resulting in a flexible noise simulator. Cheng \textit{et al.} \cite{cheng2018towards} introduced a non-parametric empirical Bayes method to optimize the multiplicative multi-scale innovation transform. Wang \textit{et al.} \cite{wang2020practical} provided a method to absorb different types of noise into a unique Gaussian noise model invariant to camera settings. These advanced noise models further improve the denoising performance of existing CNNs.

Meanwhile, some methods revisited the influence of demosaicing and tried to combine it with denoising to better capture noise statistics: Gharbi \textit{et al.} \cite{gharbi2016deep} trained a network using the noise level as an extra input, leading to a significant improvement in both demosaicing and denoising. Kokkinos \textit{et al.} \cite{kokkinos2018deep} introduced a residual network that performs these joint operations iteratively. Qian \textit{et al.} \cite{qian2019trinity} presented a multi-task model for joint demosaicing, denoising, and super-resolution, where a residual-in-residual dense block was used to obtain the optimal performance. Liu \textit{et al.} \cite{liu2020joint} further employed a density map as a self-guidance for the network to help focus more on high-frequency patterns. 

\subsection{Joint Image Enhancement and Denoising}
Since aesthetic degradation is often accompanied by obvious noise in low-light images, the above methods that target either enhancement or denoising become less optimal. Although a cascaded application of enhancement and denoising is feasible \cite{li2015low, wei2018deep, guo2019pipeline}, it is preferable to perform these operations jointly. To this end, some non-learning-based methods have been proposed with the help of handcrafted priors: Su \textit{et al.} \cite{su2017low} introduced a two-step noise suppression model, where a noise-aware histogram is first obtained to prevent noise amplification during contrast enhancement. Then a just noticeable difference model is used to suppress noise further. Ren \textit{et al.} \cite{ren2018joint} performed Retinex theory-based sequential image decomposition to obtain a smooth illumination map as well as a noise-free reflectance map that allows for further adjustment. Li \textit{et al.} \cite{li2018structure} modified the Retinex model with a noise term and proposed an optimization method that predicts illumination, reflectance, and noise map simultaneously.

The joint task is more challenging when it comes to CNNs. With the help of existing datasets \cite{wei2018deep, chen2018learning} that provide short-long exposure image pairs, a certain amount of progress has been made \cite{zhang2019kindling, zhang2020principle, zhang2020attention, xu2020learning}. Several works have exploited an alternative approach where synthetic images were used during the training stage. Wang \textit{et al.} \cite{wang2019progressive} synthesized Poisson-Gaussian noise using a simplified inverted ISP pipeline to train their progressive Retinex model. Chang \textit{et al.} \cite{chang2020low} investigated the distributions between short-exposed, long-exposed, and exposure-corrected images and proposed an improved data synthesis method for their short-long exposure fusion network. For unpaired learning, a two-stage GAN-based network was proposed in \cite{xiong2020unsupervised}, where the Retinex theory-based enhancement is performed first, followed by a denoiser trained with the help of pseudo triples. Yang \textit{et al.} \cite{yang2021lowlight} proposed a two-stage recurrent method that combines full-supervision and adversarial learning. Anantrasirichai \textit{et al.} \cite{anantrasirichai2021contextual} adapted the CycleGAN structure for enhancing high-resolution inputs via contextual colorization and denoising.  

Although these works have yielded good performance, there is still room for improvement. On the one hand, non-learning-based methods rely on handcrafted priors that do not realize the model-specific noise statistics and the high-level human perceptual aesthetics. On the other hand, most learning-based approaches only manipulate illumination and lack color enhancement. Consequently, the quality of the obtained images remains less satisfactory. In the literature, there are two works close to our work: \cite{atoum2020color} employed two networks to handle illumination and color in LAB color space. While Yang \textit{et al.} \cite{yang2020fidelity} proposed a semi-supervised network that performs band decomposition with denoising on the input image and then recomposes them to ensure the aesthetic quality is aligned with human visual perception. However, similar to other methods, they still rely on existing short-long exposure image pairs limited to specific camera models.

Compared to existing works, the elaborate two-branch structure in the proposed framework enables illumination enhancement, color enhancement, and model-specific denoising to operate jointly. To leverage the lack of paired data for multi-task learning, a weakly supervised strategy is introduced to make the training feasible. The proposed method does not require existing low-light/normal-light paired datasets for training the networks and thus can be adapted to new camera models without much effort.

\section{Proposed Method}
This section provides details of the proposed framework. We first present the overall pipeline to offer the reader a brief understanding of the workflow. Then we introduce the design of each branch in detail.

\subsection{Framework Overview}
\begin{figure*}[htb]
    \centering
    \includegraphics[scale=1]{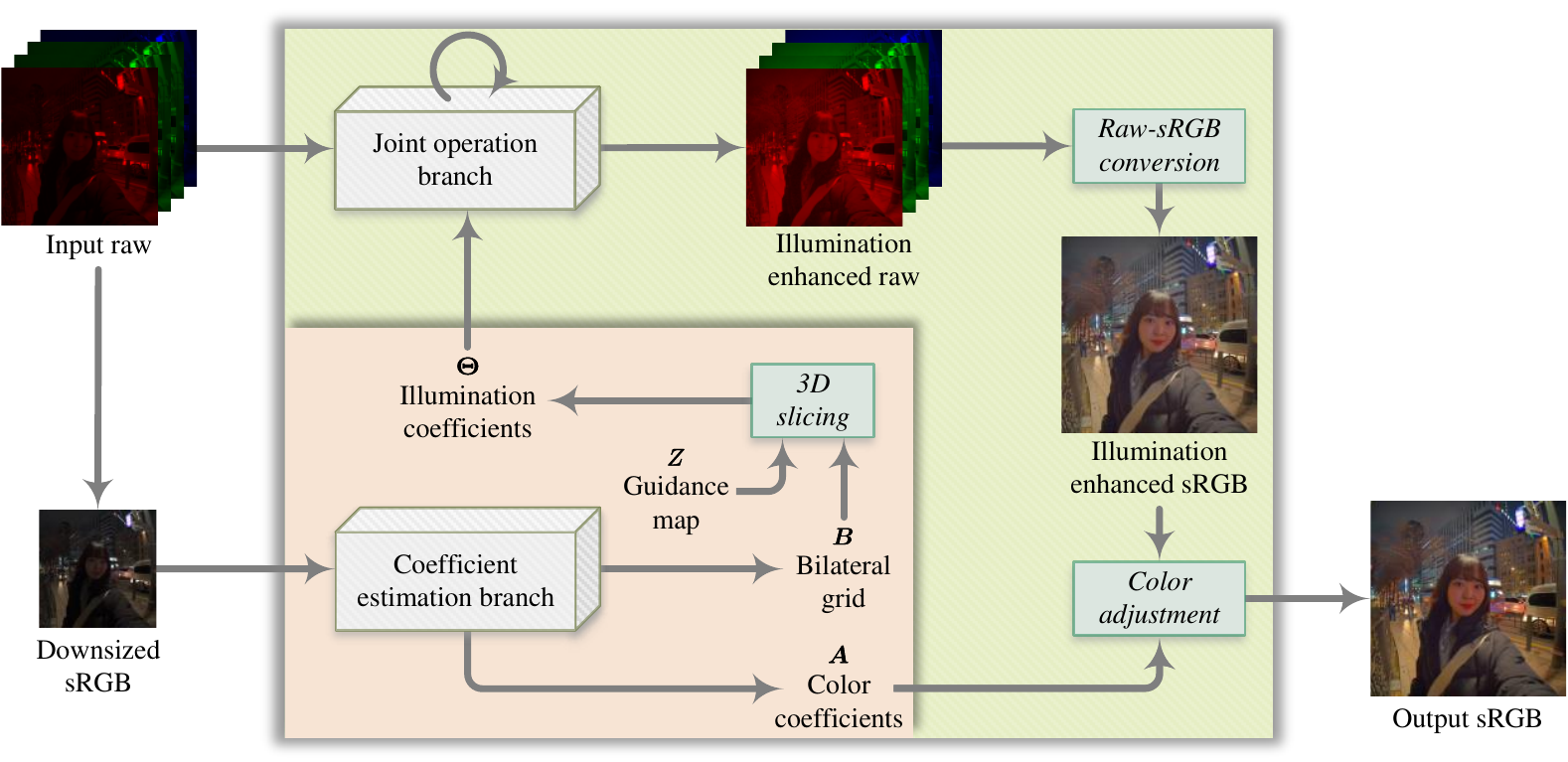}
    \caption{Overview of the proposed two-branch framework. The coefficient estimation process is grouped in light pink, while the joint enhancement and denoising process is grouped in light olive.}
    \label{fig_s3a1}
\end{figure*}

An overview of the proposed framework is shown in Fig. \ref{fig_s3a1}. It mainly consists of two branches: the coefficient estimation branch and the joint operation branch. Given an unprocessed camera raw image as the input, we first convert it to sRGB color space and resize to $256\times256$. There are two reasons for using the downsampled version instead of the original one: First, the receptive field of the network cannot cover the input images with ultra-high resolution, and thus the global information learned by the network can be incomplete; Second, the effects of noise can be significantly suppressed by downsampling. The downsampled sRGB image is then passed to the coefficient estimation branch to learn two sets of adjustment parameters: the illumination adjustment coefficients and the color adjustment coefficients. In contrast to the coefficient estimation branch, the joint operation branch processes the full-resolution raw image. In particular, an elaborate lightweight network performs joint illumination enhancement and denoising progressively, resulting in a denoised image with enhanced illumination. Next, the enhanced raw image is demosaiced and converted to an sRGB image, and a polynomial transformation is applied to improve the color and contrast. As a result, the obtained image has better brightness with less noise and looks visually more pleasing.
 
The two-branch design takes advantage of several aspects in real-world mobile imaging. On the one hand, the adjustment parameters are learned in low-resolution space, which gives the benefit of low consumption of hardware resources and fast processing speed and makes the process robust to different noise characteristics and arbitrary input resolutions. On the other hand, the enhancement and denoising are performed jointly, which is preferred as it avoids artifacts, such as amplified noise or over-smoothed details, which frequently occur in previous methods.

\begin{figure*}[htb]
    \centering
    \includegraphics[scale=1]{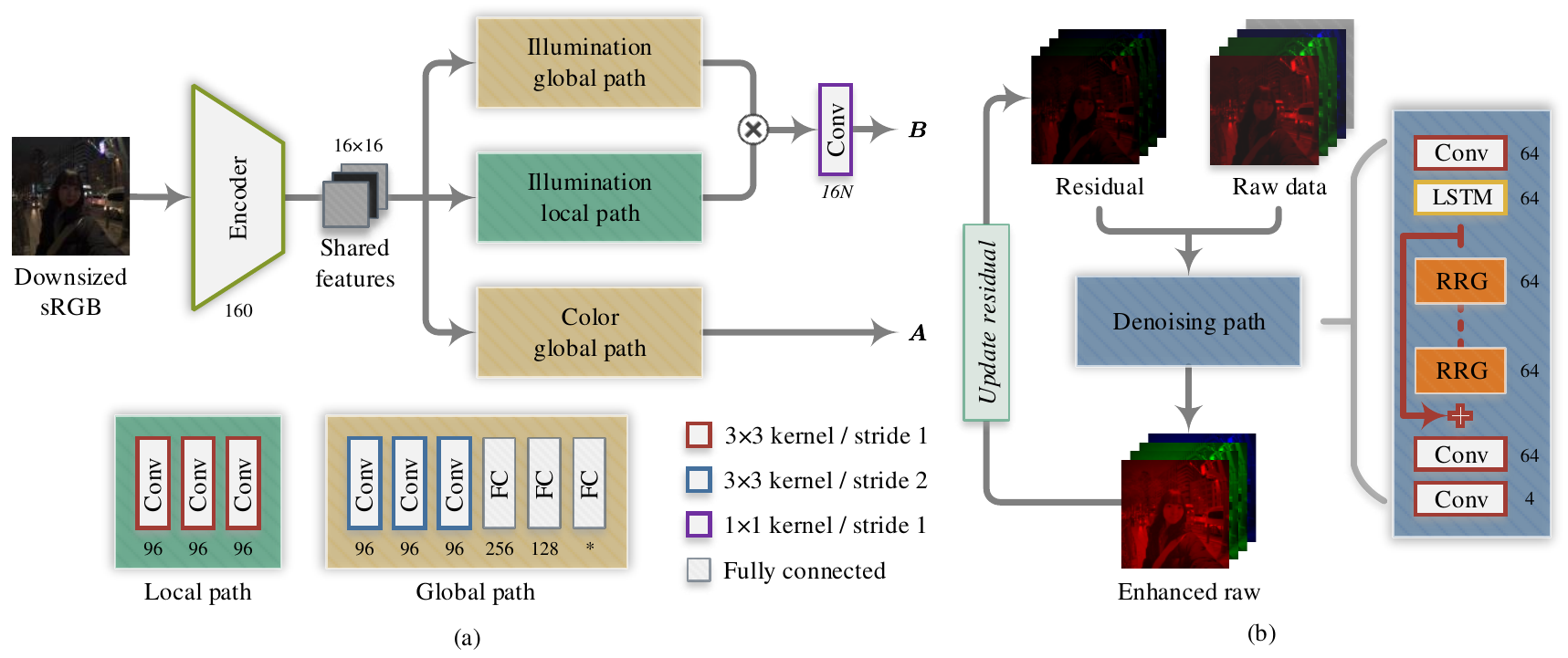}
    \caption{The network structure of (a) the coefficient estimation branch and (b) the joint operation branch. Note that the raw data in (b) include the captured raw data $I_{0}$ and its corresponding noise variance map $V$. The number of output channels is shown underneath or next to each layer.}
    \label{fig_s3bc1}
\end{figure*}

\subsection{Coefficient Estimation Branch}
The main structure of the coefficient estimation branch is depicted in Fig. \ref{fig_s3bc1}(a). The features of the downsampled sRGB image are extracted by an encoder at the early stage, here we employ MobileNet v2 \cite{sandler2018mobilenetv2} as the backbone feature extractor, in which all the batch-norm layers are removed. The sixth bottleneck module is empirically chosen as the output block because this module provides deeper features with more vital encoding ability while keeping the spatial resolution of the output feature maps relatively high ($16\times16$ for the input size of $256\times256$). The extracted features are then shared to predict the enhancement coefficients.

We observe that a human expert usually starts with global operations followed by local adjustments for illumination enhancement. Inspired by \cite{gharbi2017deep}, we design two paths, namely global path and local path, to learn the features of global and local illumination adjustments, respectively. Since these adjustments do not work well when applied separately (see Section IV-B for more details), we implicitly combine them in the feature space and use a convolution layer to map the fused features to the desired dimensions. The output is a set of bilateral grids, \textit{i.e.}, ${\textbf{\textit{B}}} = \left\{ {{B_1},{B_2}, \cdot \cdot \cdot ,{B_N}} \right\}$, where $N$ is the number of total iterations. Each ${B_n}$ has the size of $16\times16\times16$. The full-resolution coefficient map $\Theta_{n}$ is then obtained by the learned bilateral upsampling via 3D slicing as follows:

\begin{equation}
    \Theta_{n}\left( x,y \right) = \sum_{i,j,k} \tau\left( x,i \right) \tau\left( y,j \right) \tau\left( z_{x,y}, k \right) B_{n}\left( i,j,k \right),
\end{equation}

\noindent where $\tau$ is the linear interpolation kernel. $(x,y)$ and $(i,j,k)$ represent the pixel coordinates and the cell index of the bilateral grid, respectively. $z_{x,y}$ is estimated from the pixel values at $(x,y)$, which will be explained in Section III-D. Note that all the coefficients for every iteration are obtained before the joint enhancement and denoising process.

In contrast to illumination adjustment, color retouching is often performed globally. Therefore, we design a color path that extracts global features and predicts a set of polynomial coefficients $\boldsymbol{A}_K$ for the color enhancement, which will be explained next.

\subsection{Progressive Enhancement without Denoising}
Since the effort required in low-light image enhancement varies a lot due to heavily fluctuated illumination distributions, the network should fit a highly diverse model to achieve dynamic local adjustments. However, learning such complicated curves could be challenging in a single forward operation. A more flexible alternative is to approximate these higher-order functions using simpler representations via recursion. The benefit is twofold: First, the image quality can be progressively improved during iterations; Second, denoising can be integrated together as a joint operation, which will be presented in Section III-D.

For clarification purposes, we first introduce the progressive enhancement using $\Theta_{n}$ and $\boldsymbol{A}_K$ without considering denoising. It begins with progressive illumination adjustment that compensates for insufficient lighting in dark regions. Motivated by \cite{guo2020zero}, the illumination adjustment at each iteration is performed through a transform function defined as follows:

\begin{equation}
    I_{n}\left( {\bf{x}} \right) = I_{n-1}\left( {\bf{x}} \right) + \Theta_{n}\left( {\bf{x}} \right) \left( 1 - L_{n-1} \left( {\bf{x}} \right) \right) I_{n-1}\left( {\bf{x}} \right),
\end{equation}

\noindent where $I_{n}$ and $I_{n-1}$ are the images in camera raw color space from the current and the previous iterations, respectively. $L_{n-1}$ is the luminance channel of $I_{n-1}$ in XYZ color space. ${\bf{x}}$ denotes a pixel coordinate vector, which will be omitted hereafter when unnecessary.

The image enhanced by the above step has improved illumination and details compared to the original input but still lacks adequate contrast and perceptually pleasing color. To cope with this issue, color enhancement is applied to improve the aesthetic quality further. We adapt the polynomial transform \cite{hong2001study} as the color enhancement operation, defined as follows:

\begin{equation}
J_{O}\left( {\bf{x}} \right) = {\boldsymbol{A}_K}{\boldsymbol{\rho}_K}\left( {J_N\left( {\bf{x}} \right)} \right),
\end{equation}

\noindent where $J_N$ represents the sRGB image converted from $I_N$, and $J_{O}$ is the color-enhanced sRGB image. $\boldsymbol{\rho}_{K}\left( {J_N\left( {\bf{x}} \right)} \right)$ is a column vector consisting of some polynomial terms of ${J_N\left( {\bf{x}} \right)}$ with the degree up to $K$, defined as follows:

\begin{equation}
    {\boldsymbol{\rho} _K}\!\left( {J_N\!\left( {\bf{x}} \right)} \right) \!= \!\left[ {r^ig^jb^k \!\mid\! {J_N\!\left( {\bf{x}} \right)}\!=\!(r,g,b), i\!+\!j\!+\!k \!\leq\! K} \right],
\end{equation}

\noindent where $i$, $j$, and $k$ are non-negative integers.

\subsection{Joint Operation Branch}
As low-light images always come with noise, one would seek a cascaded design composing an enhancement network and a denoising network to tackle this issue. This is, however, not an optimal design. On the one hand, if denoising is performed first, even though the remaining noise can be less noticeable, its visibility could be significantly increased after enhancement. In other words, it results in increased (and possibly inconsistent) dispersion of the pixel values. On the other hand, if enhancement is performed first, regions with similar illumination in the resultant image will likely contain various noise levels. This might lead to more aggressive denoising that will produce over-smoothed details. Based on these observations, we introduce a joint operation branch that achieves a good balance between the two operations.

Let $R_{n} = \Theta_{n} (1-L_{n-1}) I_{n-1}$, then (2) can be rewritten as follows:

\begin{equation}
    \begin{aligned}
        I_{N} &= I_{N-1} + R_{N}\\
        &= I_{N-2} + R_{N-1} + R_{N}\\
        &= I_{0} + R_{1} + R_{2} + \cdots + R_{N-1} + R_{N}.
    \end{aligned}
\end{equation}
Apparently, the enhancement result $I_{N}$ is the sum of the input $I_{0}$ and all the residuals obtained from each iteration. Hence, a residual network $\mathcal{G}$ can be introduced to convert the two operations to a single one as follows:

\begin{equation}
    I_{N} = \mathcal{G} \left( I_{0}, \sum_{n=1}^{N} R_{n} \right).
\end{equation}
Although such a network is intuitively straightforward, it does not work well for two reasons: First, the noise statistics can be altered at each iteration, which makes the corresponding noise modeling hard to be followed; Second, the sum of all the residuals contains accumulated noise, which pushes the denoising to be more aggressive in order to suppress these noise. These problems can be alleviated if denoising is embedded into each iteration with the network being able to update the noise features dynamically. To this end, we insert a long short-term memory (LSTM) layer with a hidden state $h$ into $\mathcal{G}$ to convert it to a recurrent residual network $\mathcal{G'}$. The update at the $n$-th iteration is thus presented as follows:

\begin{equation}
    \left\{ \begin{array}{l}
    R_{n} = \Theta_{n}\left( 1-L^U_{n-1} \right)I^U_{n-1},\\
    \left\{ {I^U_{n}, h_{n}} \right\} = \mathcal{G'} \left( I_{0}, V, R_{n}, h_{n-1} \right),
    \end{array} \right.
\end{equation}

\noindent where $I^U_{n}$ denotes the updated image, and $V$ is the estimated pixel-wise noise variance map. For the first iteration, $I^U_{0}$ is obtained as follows:

\begin{equation}
    \left\{ {I^U_{0}, h_{0}} \right\} = \mathcal{G'} \left( I_{0}, V, \mathbf{0}, \mathbf{0} \right).
\end{equation}
By iteratively applying (7), the joint enhancement and denoising can be performed progressively. This design not only produces more consistent denoising results since it takes both the raw input and the enhancement residual into account but also preserves more details after denoising as the enhancement residual at each iteration is relatively moderate.

We adapt the iterative ResNet \cite{knaus2014progressive} and CycleISP \cite{zamir2020cycleisp} in the network design, where the structure is shown in Fig. \ref{fig_s3bc1}(b) and the recursive residual group (RRG) here is composed of several dual attention blocks (DABs). From $I^U_{0}$, we also obtain the full-size guidance map $Z$ that contains $z_{x,y}$ values needed in (1) as follows:

\begin{equation}
    Z = \mathcal{F} \left( I^U_{0} \right), 
\end{equation}

\noindent where $\mathcal{F}$ is a mapping function learned by a simple network consisting of three convolutional layers. 

Some operations in the ISP pipeline can also contaminate the noise profile. Hence, color space conversion is performed after the joint enhancement and denoising, resulting in an sRGB image $J^{U}_{N}$. Note that our goal is to learn perceptual enhancement rather than color correction; therefore, we assume the white point from camera metadata is correct and directly use it in the color space conversion. Finally, by replacing $J_{N}$ with $J^{U}_{N}$ in (3), we obtain the result with higher aesthetic quality.

\subsection{Loss Function}
In general, training a neural network requires massive paired data. However, to the best of our knowledge, no dataset in the literature contains low-light images of real-world scenes and their corresponding noise-free ground-truth images with expert enhanced illumination and color. In practice, making such a dataset is very challenging. Therefore, we design a strategy that trains the coefficient estimation branch in a weakly supervised manner and the joint operation branch in a supervised manner using synthetic paired samples.

We employ three loss terms to train the coefficient estimation branch: exposure loss, perceptual loss, and color loss. First, we adapt the weighting function from \cite{tom2007exposure} as the exposure loss for encouraging the network to adjust both underexposed and overexposed regions to an adequate exposure range. The exposure loss is defined as follows:

\begin{equation}
    \mathscr{L}_{e} = \frac{1}{M} \sum_{\bf{x}} \left( 1 - \exp \left( -\frac{\left( I_{N}({\bf{x}}) - 0.5 \right)^2}{2\delta^2} \right ) \right),
\end{equation}

\noindent where $M$ is the total number of pixels, and $\delta$ is a hyper-parameter that controls the aesthetic quality.

Although the exposure loss effectively enhances illumination, it results in poor contrast if used alone. To preserve the overall perceptual quality, based on the observation that VGG models are not very sensitive to illumination changes \cite{jiang2019enlightengan}, the Euclidean distance of VGG16 features is used as the perceptual loss to ensure that the input and output images have similar feature presentations:

\begin{equation}
    \mathscr{L}_{p} = \frac{1}{5} \sum_{i=1}^{5} \left\| \mathcal{V}_{i} \left( I_{N} \right) - \mathcal{V}_{i} \left( I_{0} \right) \right\|_{2},
\end{equation}

\noindent where $\mathcal{V}_{i}$ represents the output features of the $i$-th block in the pre-trained VGG16 model \cite{simonyan2014very}.

\begin{figure}[htb]
    \centering
    \includegraphics[scale=1]{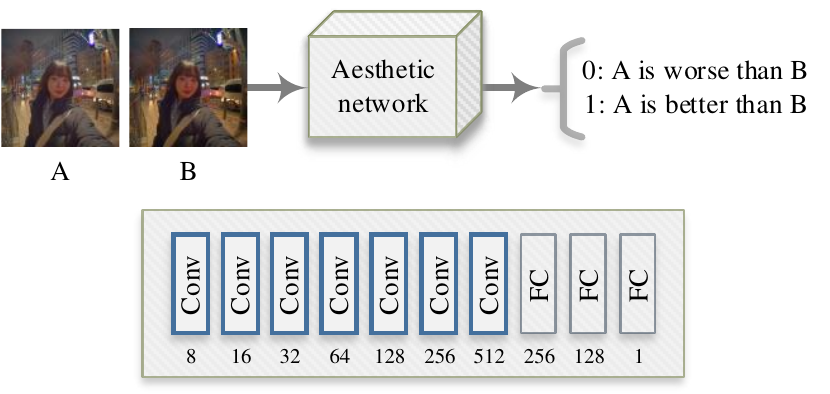}
    \caption{Structure of the aesthetic network.}
    \label{fig_s3e1}
\end{figure}

The color loss is more challenging because it is not easy to quantify the aesthetic quality of an image. Motivated by \cite{talebi2018learned} that uses a pre-trained image assessment network as a part of the supervision, we employ an aesthetic network $\mathcal{Q}$ that evaluates the aesthetic quality. However, we experienced difficulties training the aesthetic network from only a single image. We notice that the original untouched image can be used as a reference in the evaluation. Therefore, we adapt the relativistic classifier in \cite{kim2019grdn} as the aesthetic network. As shown in Fig. \ref{fig_s3e1}, it uses paired images obtained before and after the color enhancement operation as ordered inputs and predicts a binary label indicating whether the enhanced result has ``better'' (1) or ``worse'' (0) aesthetic quality. The color loss is defined as follows:

\begin{equation}
    \begin{aligned}
        \mathscr{L}_{c} &= \hat{y} \left( -\log \left( \mathcal{Q} \left( J_{N}, J_{O} \right) \right) \right)\\
        &+ \left(1 - \hat{y} \right) \left( -\log \left( 1 - \mathcal{Q} \left( J_{O}, J_{N} \right) \right) \right),
    \end{aligned}
\end{equation}

\noindent where $\hat{y}$ represents the ground-truth binary label. The final loss function for training the coefficient estimation branch is the combination of the three terms:

\begin{equation}
    \mathscr{L}_{E}= \mathscr{L}_{e} + \mathscr{L}_{p} + \mathscr{L}_{c}.
\end{equation}
Note that the aesthetic network is not trained with other branches. Instead, we train it separately using paired data and kept it fixed, likewise with the pre-trained VGG16 model.

To train the joint operation branch, we freeze the coefficient estimation branch and compute the loss for each iteration as follows:

\begin{equation}
    \mathscr{L}_{D} = \frac{1}{N+1} \sum_{n=0}^{N} \left\| I_{n}^{U} - I_{n} \right\|_{1},
\end{equation}

\noindent where $I_{n}^{U}$ and $I_n$ are obtained from (7) and (2) using noisy and clean raw images, respectively. As a result, the joint operation branch can learn how to denoise raw images during iterations.

\subsection{Noise Synthesis}
The joint operation branch works in raw domain where the noise statistics are device-dependent. Hence, retraining of this branch is necessary for new camera models. To obtain sufficient data for retraining, a straightforward approach is to collect burst images using the given camera and then fuse them to make ground-truth images \cite{abdelhamed2018high}. This approach is very time-consuming and sensitive to the misalignment of the captured images. A more convenient method is to synthesize realistic noisy-clean image pairs. Specifically, our method requires a clean dataset for all camera models, where we used the raw images from the RAISE dataset \cite{dang2015raise}. Given a new camera model, we only estimate noise statistics and generate noisy-clean pairs by adding the camera-specific noise to clean raw images. Thus, there is no need to recollect paired images for new camera models. 

To generate realistic device-dependent noise, we pay attention to two major sources of noise: photon noise (dubbed $N_{p}$) from the quantum characteristics of photons and circuit noise from the camera circuitry. Although safe shutter speed effectively mitigates camera motion blur, the captured images are inevitably darker than those taken with long exposure. Since the perturbation of circuit noise becomes more evident in low-light conditions, instead of utilizing the commonly used heteroscedastic Gaussian noise model, we adapt a recent work \cite{wei2020physics} that characterizes the long-tail mixture of dark current noise, thermal noise, and source follower noise by a Tukey lambda distribution (dubbed $N_{r}$), and combine horizontal banding pattern noise (dubbed $N_{b}$) and quantization noise (dubbed $N_{q}$) as the circuit noise model. Their probability distributions are given as follows:

\begin{equation}
    \begin{aligned}
        \left( E + N_{p} \right) &\sim \mathcal{P} \left( E \right),\\
        N_{r} &\sim TL \left( \lambda_{r};0,\sigma_{r} \right),\\
        N_{b} &\sim \mathcal{N} \left( 0,\sigma_{b} \right),\\
        N_{q} &\sim \mathcal{U} \left( -0.5s,0.5s \right),
    \end{aligned}
\end{equation}

\noindent where $E$ is the number of photons and $TL$ represents the Tukey lambda distribution with the shape determined by $\lambda_{r}$. $\mathcal{P}$, $\mathcal{N}$, and $\mathcal{U}$ represent the Poisson, Gaussian, and uniform distributions, respectively. $s$ is the quantization step.

The parameters of circuit noise can be estimated from lightless samples\footnote{See \cite{wei2020physics} for more details.}: $\sigma_{b}$ is estimated using the mean values of each row from the raw images; $\lambda_{r}$ and $\sigma_{r}$ are estimated using the probability plot correlation coefficient plot and probability plot, respectively. The final noise model $N_{f}$ is defined as the combination of all the above noise sources:

\begin{equation}
    N_{f} = \kappa N_{p} + N_{r} + N_{b} + N_{q},
\end{equation}

\noindent where $\kappa$ is the conversion gain. It can be derived from flat-field frames via the photon transfer method \cite{janesick1985ccd} or extracted from camera calibration metadata if available.

\section{Experiment Results}
This section presents the experiment results and analysis. Notably, each branch was first evaluated individually and then compared with its related state-of-the-arts. Then the whole framework was evaluated to reveal its promising potential. All our results reported are with color enhancement unless otherwise mentioned.

\subsection{Implementation Details}
We found that the bilateral upsampling and the weak supervision raise the difficulty in training the coefficient estimation branch from scratch. We thus performed pre-training to handle this issue, where we generated linear and non-linear sRGB image pairs as training data. The pre-training encourages the network to learn an initial mapping between linear to non-linear sRGB images, which was effective for the convergence of subsequent training. The ``Expert C'' subset of the Adobe FiveK dataset \cite{fivek} was used to train the aesthetic network. We modified the input images to the same exposure levels as the retouched ones to avoid biased learning due to exposure differences. After that, we fixed the aesthetic network and trained only the coefficient estimation branch. For this fine-tuning, training images were chosen from the clean raw subset (obtained in burst mode) of the HDR+ dataset \cite{hasinoff2016burst}. We used the exact input resolution of $256\times256$ in both the pre-training and the fine-tuning to avoid image resizing.

For the joint operation branch, we selected the ``Outdoor'' subset of the RAISE dataset \cite{dang2015raise} as the training set. To generate degraded images, the original raw images were first darkened by random scale factors in the range of 1-16, which simulates exposure compensation of up to 4 stops, then cropped to $64\times64$ and mixed with the synthetic noise.

The proposed framework was implemented in PyTorch and trained on an Nvidia Titan RTX GPU\footnote{Code available at https://github.com/YCL92/LowlightENR}.

\subsection{Component Analysis}
The proposed framework performs three operations: illumination adjustment, color enhancement, and denoising. We conducted three experiments to find out the optimal configuration for these operations.

\begin{figure}[htb]
    \centering
    \begin{tikzpicture}[scale=0.8]
        \begin{axis}[xlabel={iteration}, ylabel={PSNR (dB)}]
        \addplot+[sharp plot] 
        coordinates
        {(2, 23.3759) (3, 24.0804) (4, 24.7890) (5, 24.8998) (6, 24.9595) (7, 25.0910) (8, 25.3178) (9, 25.3492) (10, 25.2130) (11, 25.3024) (12, 25.2446) (13, 25.2306) (14, 25.1312) (15, 25.3502) (16, 25.2863)};
        \end{axis}
    \end{tikzpicture}
    \caption{Illumination adjustment results of different numbers of iterations.}
    \label{fig_s4b1}
\end{figure}
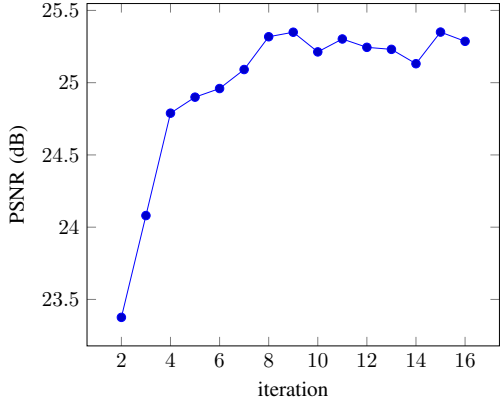

First, we examined the performance of the illumination adjustment. According to Section III-C, it is expected that the image quality keeps improving with more iterations until reaching the network's capacity. Therefore, we applied (2) with different numbers of iterations ranging from 2 to 16 and trained each variant on a modified version of the Adobe FiveK dataset without color shift. The results are depicted in Fig. \ref{fig_s4b1}. It can be seen that the average PSNR gradually increases for about 9 iterations and then starts to fluctuate. Based on this observation, we set $N=9$ as the default.

Second, we examined the performance of the color adjustment. To this end, we selected eight types of polynomial transforms with various degrees and trained each variant on a modified version of the Adobe FiveK dataset without exposure shift. The first four candidates, $\boldsymbol{\rho}_1$-$\boldsymbol{\rho}_4$, are expressed as follows:

\begin{equation}
    \left\{
    \begin{aligned}
        \boldsymbol{\rho}_1 = [ &r,g,b,1 ]^T,\\
        \boldsymbol{\rho}_2 = [ &r^2,g^2,b^2,rg,gb,rb,r,g,b,1 ]^T,\\
        \boldsymbol{\rho}_3 = [ &r^3,g^3,b^3,rg^2,gb^2, rb^2,gr^2,bg^2,br^2,rgb, \\
        &r^2,g^2,b^2,rg,gb,rb,r,g,b,1 ]^T,\\
        \boldsymbol{\rho}_4 = [ &r^4,g^4,b^4,r^3g,r^3b,g^3r,g^3b,b^3r,b^3g,\\
        &r^2g^2,g^2b^2,r^2b^2,r^2gb,g^2rb,b^2rg,\\
        &r^3,g^3,b^3,rg^2,gb^2,rb^2,gr^2,bg^2,br^2,rgb,\\
        &r^2,g^2,b^2,rg,gb,rb,r,g,b,1 ]^T.
    \end{aligned}
    \right.
\end{equation}

\noindent The remaining four candidates, $\overline{\boldsymbol{\rho}}_1$-$\overline{\boldsymbol{\rho}}_4$, are obtained by excluding the constant term ``$1$'' from $\boldsymbol{\rho}_1$-$\boldsymbol{\rho}_4$, respectively. As shown in Table \ref{table_s4b1}, the highest degree $\boldsymbol{\rho}_4$ resulted in the best performance but with significant computation cost, and we thus chose $\boldsymbol{\rho}_3$ as the default.

\begin{table}[htb]
    \renewcommand{\arraystretch}{1.3}
    \caption{Color Adjustment Results of Different Polynomial Transforms}
    \label{table_s4b1}
    \centering
    \begin{tabular}{ccccc}
        \hline
        Polynomial & $\overline{\boldsymbol{\rho}}_1$ & $\overline{\boldsymbol{\rho}}_2$ & $\overline{\boldsymbol{\rho}}_3$ & $\overline{\boldsymbol{\rho}}_4$\\
        PSNR & 19.1429 & 24.6043 & 28.2288 & 28.3260\\
        \hline
        \hline
        Polynomial & $\boldsymbol{\rho}_1$ & $\boldsymbol{\rho}_2$ & $\boldsymbol{\rho}_3$ & $\boldsymbol{\rho}_4$\\
        PSNR & 26.6424 & 31.6538 & 33.7718 & 34.9111\\
        \hline
    \end{tabular}
\end{table}

Third, we tested the performance of the joint operation branch. To this end, we trained a series of variants with different numbers of DAB (dubbed $n_{d}$) and RRG (dubbed $n_{r}$) using the SIDD dataset \cite{abdelhamed2018high}. The enhancement coefficients were estimated from the clean images and fixed for all network variants. To reduce computation complexity, we limited the number of convolution layers so that $n_{r}n_{d} \leq 16$. From the results shown in Table \ref{table_s4b2}, we found that a deeper network indeed helps improve the performance at the expense of computation cost, and we set $n_{r}=6$ and $n_{d}=2$ considering the trade-off between accuracy and computation complexity.

\begin{table}[htb]
    \begingroup
    \setlength{\tabcolsep}{5pt}
    \renewcommand{\arraystretch}{1.3}
    \caption{Denoising Results of Different Denoising Network Configurations in Terms of PSNR}
    \label{table_s4b2}
    \centering
    \begin{tabular}{ccccccc}
        \hline
        $n_{r}$/$n_{d}$ & 2/3 & 3/2 & 2/4 & 4/2 & 2/5 & 5/2\\
        PSNR & 39.9417 & 39.9964 & 40.1302 & 40.0832 & 40.1827 & 40.4202\\
        \hline
        \hline
        $n_{r}$/$n_{d}$ & 2/6 & 6/2 & 2/7 & 7/2 & 2/8 & 8/2\\
        PSNR & 40.2359 & 40.7132 & 40.3649 & 40.7388 & 40.5656 & 40.8680\\
        \hline
        \hline
        $n_{r}$/$n_{d}$ & 3/3 & 3/4 & 4/3 & 3/5 & 5/3 & 4/4\\
        PSNR & 40.2093 & 40.4257 & 40.4020 & 40.3981 & 40.6784 & 40.6167\\
        \hline
    \end{tabular}
    \endgroup
\end{table}

\begin{figure}[htb]
    \centering
    \includegraphics[scale=1]{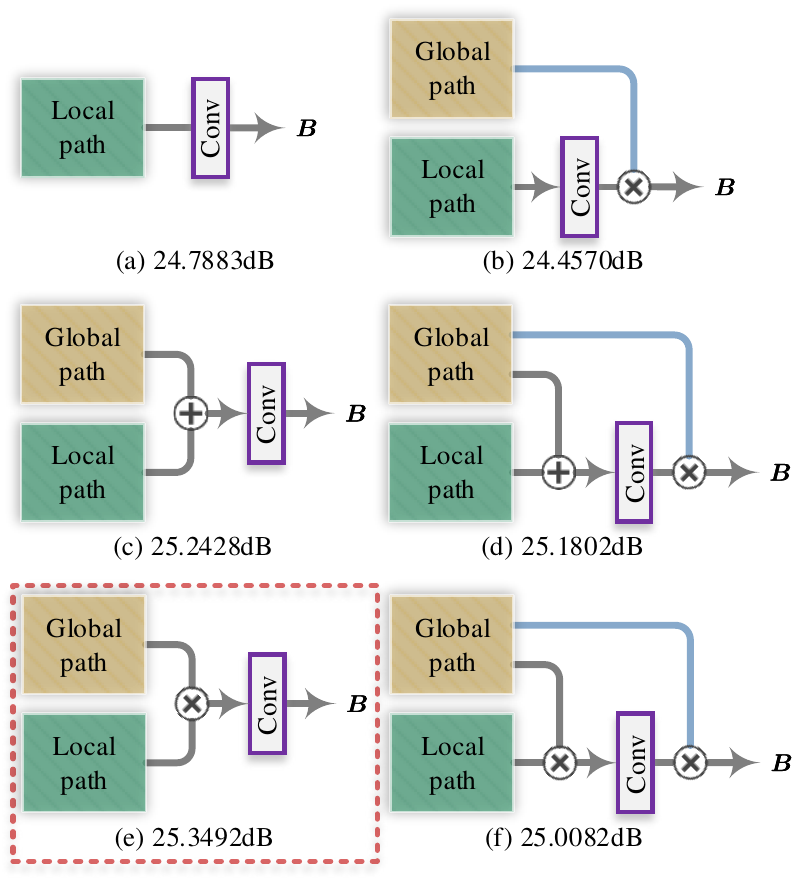}
    \caption{Variants of the illumination coefficient estimation path and their corresponding average PSNRs on the Adobe FiveK validation dataset. Note that the blue line indicates per-iteration multiplication.}
    \label{fig_s4b2}
\end{figure}

\begin{figure}[htb]
    \centering
    \includegraphics[scale=1]{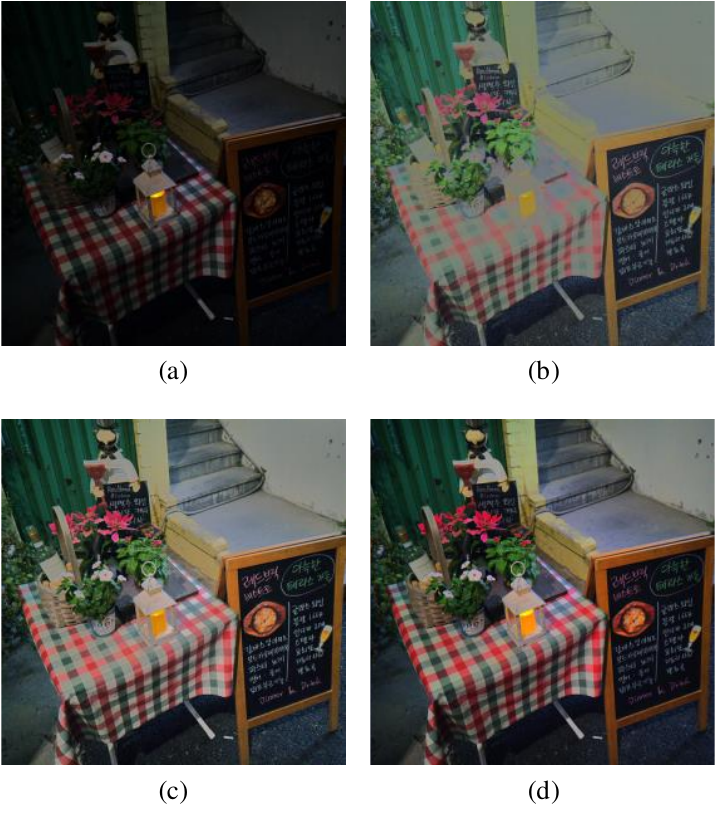}
    \caption{Enhancement results trained with different loss function variants: (a) without $\mathscr{L}_{e}$; (b) without $\mathscr{L}_{p}$; (c) without $\mathscr{L}_{c}$; (d) with all $\mathscr{L}_{e}$, $\mathscr{L}_{p}$, and $\mathscr{L}_{c}$.}
    \label{fig_s4b3}
\end{figure}

We then evaluated the individual components in each branch. For the joint operation branch, we found that removing the LSTM module leads to significant performance deterioration from $40.7132$dB to $34.5686$dB, which aligns with our analysis in Section III-D. For the coefficient estimation branch, several variants as shown in Fig. \ref{fig_s4b2} were tested. It can be seen that embedding global operations implicitly in the feature space helps learn a better understanding of the overall context, enabling further image quality improvement. 
In addition to the network structure, the loss function was also tested. Fig. \ref{fig_s4b3} shows the results obtained by the networks trained with different loss function variants. Apparently, each component in the proposed loss function plays an essential role in network training.

\begin{figure*}[b]
    \centering
    \includegraphics[scale=1]{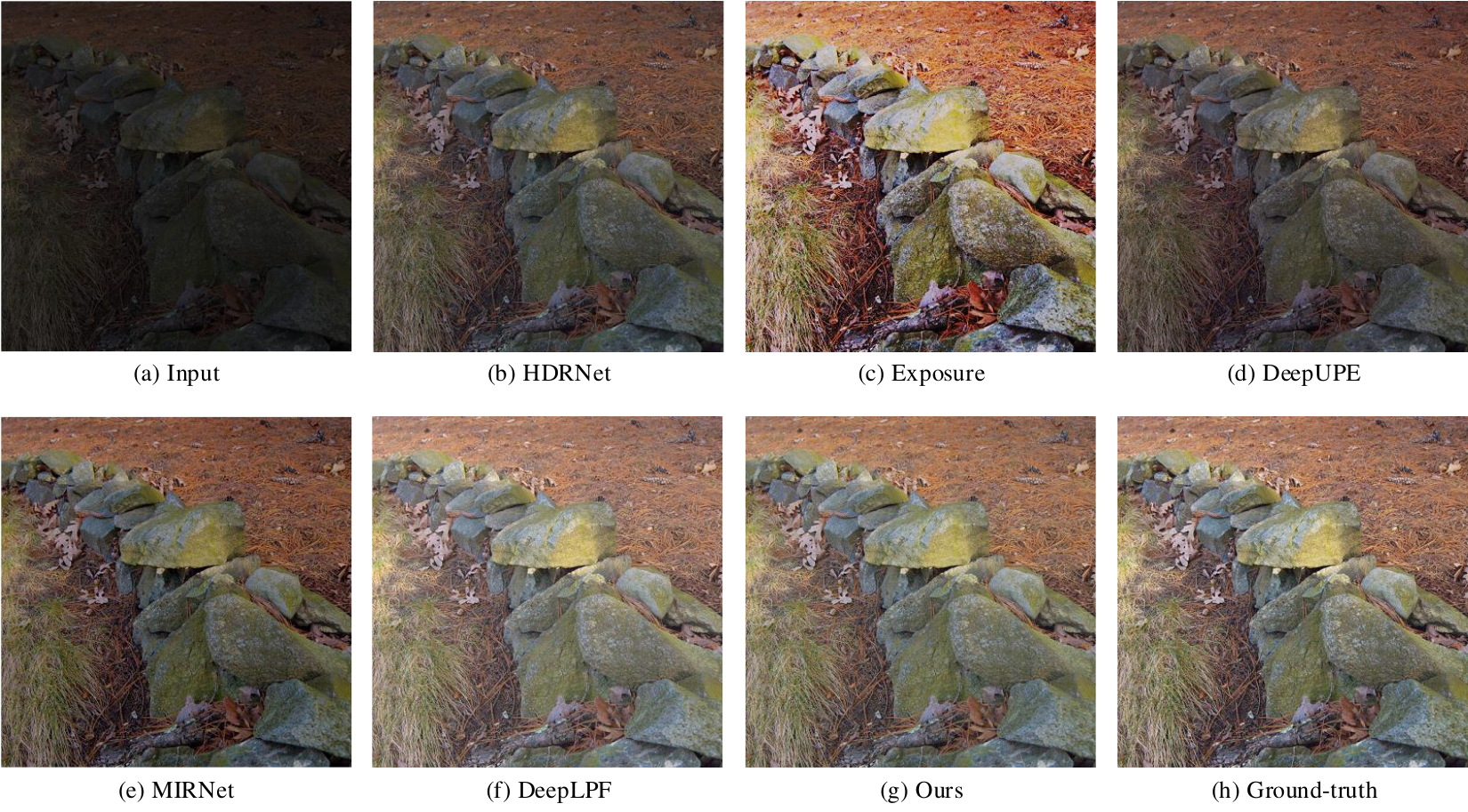}
    \caption{An example of the visual comparison of ours and several state-of-the-art image enhancement methods on the Adobe FiveK dataset.}
    \label{fig_s4c1}
\end{figure*}

\begin{table}[b]
    \renewcommand{\arraystretch}{1.3}
    \caption{Enhancement Results on The Adobe FiveK Dataset}
    \label{table_s4c1}
    \centering
    \begin{tabular}{ccccc}
        \hline
        \multirow{2}{*}{Method} & \multicolumn{2}{c}{PSNR} & \multicolumn{2}{c}{SSIM}\\
        & mean & std & mean & std\\
        \hline
        HDRNet \cite{gharbi2017deep} & 23.3912 & 5.4072 & 0.9290 & 0.0568\\
        Exposure \cite{hu2018exposure} & 18.4698 & \textbf{3.6172} & 0.7649 & 0.0971\\
        DeepUPE \cite{wang2019underexposed} & 23.3498 & 4.5180 & 0.9129 & 0.0555\\
        MIRNet \cite{zamir2020learning} & 22.2512 & 4.3131 & 0.9028 & 0.0674\\
        DeepLPF \cite{moran2020deeplpf} & 24.8260 & 5.4515 & 0.9392 & 0.0612\\
        Ours & \textbf{25.8083} & 5.2333 & \textbf{0.9508} & \textbf{0.0530}\\
        \hline
    \end{tabular}
\end{table}

\subsection{Evaluation of Coefficient Estimation Branch}
To demonstrate the learning ability of the proposed coefficient estimation branch, we first trained it under full supervision using paired data for 300 epochs with L2 loss. For the performance comparison, five state-of-the-art image enhancement methods that use paired datasets for training were selected, namely HDRNet \cite{gharbi2017deep}, Exposure \cite{hu2018exposure}, DeepUPE \cite{wang2019underexposed}, MIRNet \cite{zamir2020learning}, and DeepLPF \cite{moran2020deeplpf}. We followed the same dataset partition protocol described in \cite{wang2019underexposed} for retraining \cite{gharbi2017deep, hu2018exposure, moran2020deeplpf} using publicly available code with default settings and directly used the pre-trained weights of \cite{zamir2020learning, wang2019underexposed}. The ``Expert C'' subset of the Adobe FiveK dataset was used for both training and testing. After that, we applied all models to full-resolution test images (1-16 megapixels) and computed PSNR and SSIM as the evaluation metrics. Fig. \ref{fig_s4c1} shows an example of the visual comparison results. It can be seen that our method successfully improved the quality of the input image so that the resultant image looks more natural and close to the ground-truth. This can be further proved by the quantitative evaluation results listed in Table \ref{table_s4c1}. The proposed method outperforms all the other state-of-the-art methods, indicating that it can learn a more precise model of expert's complicated retouching effects.

\begin{table}[htb]
    \renewcommand{\arraystretch}{1.3}
    \caption{Enhancement Results on The SID Dataset}
    \label{table_s4c2}
    \centering
    \begin{threeparttable}[b]
    \begin{tabular}{ccccccc}
        \hline
        Method & PSNR & SSIM & IE & GCF & NIQMC\\
        \hline
        LIME \cite{guo2016lime} & 16.8685 & 0.4630 & 6.2399 & 4.1678 & 3.6991\\
        JED \cite{guo2020zero} & 13.0279 & 0.3370 & 3.9429 & 0.7070 & 1.0918\\
        LR3M \cite{ren2020lr3m} & 12.3882 & 0.2578 & 3.8935 & 0.6433 & 0.8926\\
        AutoExp \cite{zhang2019dual} & 14.4591 & 0.3798 & 5.5633 & 2.5777 & 2.8736\\
        Enlighten \cite{jiang2019enlightengan} & 14.8203 & 0.4503 & 4.9853 & 1.9282 & 2.4124\\
        ZeroDCE \cite{guo2020zero} & 12.6766 & 0.2243 & 4.6076 & 1.3056 & 1.8566\\
        Ours (w/o CE$^*$) & \textbf{18.2517} & \textbf{0.5098} & 6.5087 & 4.6893 & 4.1004\\
        Ours & 18.0275 & 0.4998 & \textbf{6.6182} & \textbf{4.9421} & \textbf{4.2210}\\
        \hline
    \end{tabular}
    \begin{tablenotes}
        \footnotesize
        \item[$*$] CE: color enhancement
    \end{tablenotes}
    \end{threeparttable}
\end{table}

\begin{figure*}[b]
    \centering
    \includegraphics[scale=1]{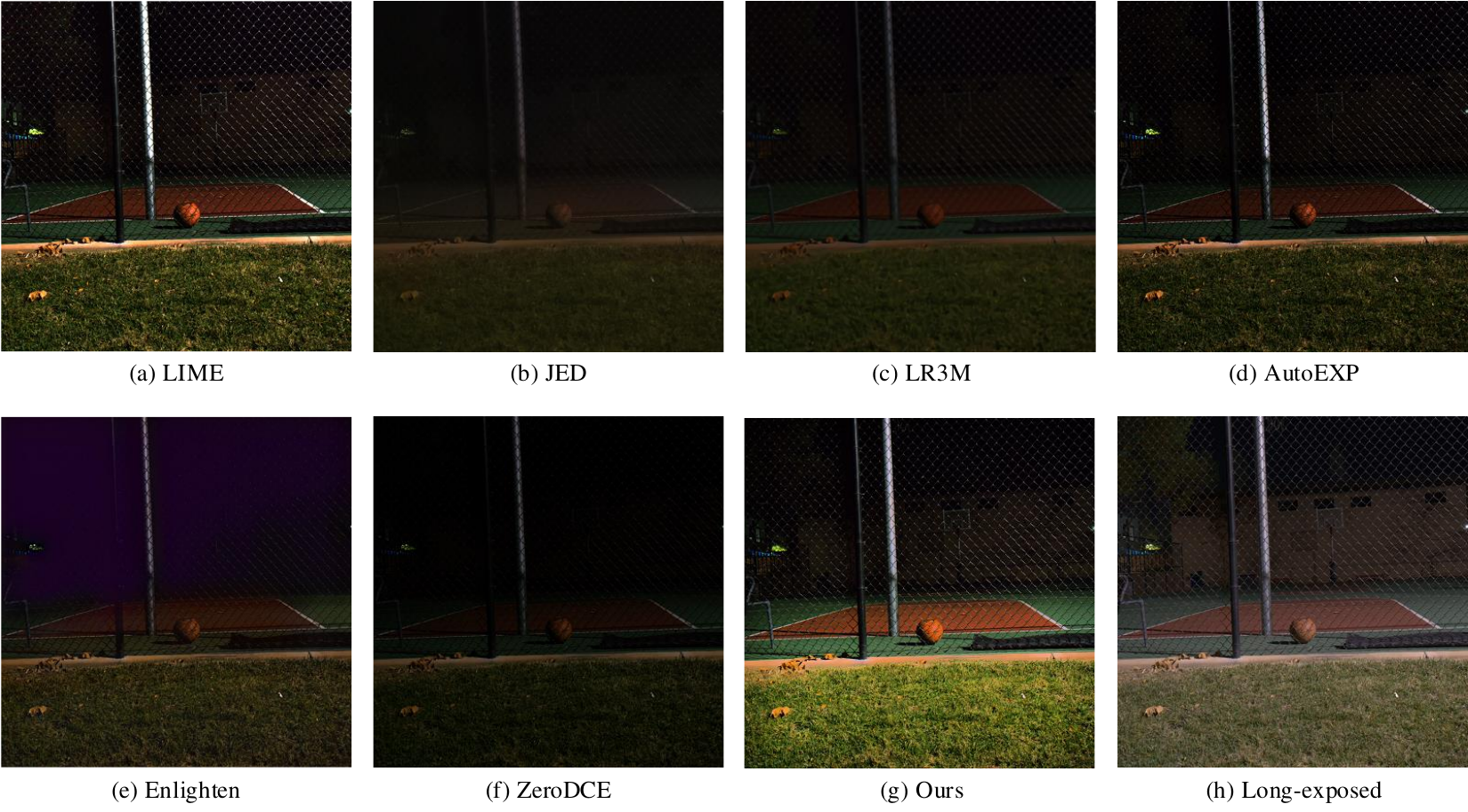}
    \caption{An example of the visual comparison of ours and several state-of-the-art image enhancement methods on the SID dataset.}
    \label{fig_s4c2}
\end{figure*}

Another experiment was performed to evaluate the coefficient estimation branch under weak supervision. To this end, we selected six handcrafted or unpaired-learning-based image enhancement methods, namely LIME \cite{guo2016lime}, JED \cite{ren2018joint}, LR3M \cite{ren2020lr3m}, AutoExp \cite{zhang2019dual}, Enlighten \cite{jiang2019enlightengan}, and ZeroDCE \cite{guo2020zero}, and directly applied them to the full-size short exposure set of the SID dataset \cite{chen2018learning} which was unseen during the training. From Fig. \ref{fig_s4c2}, it can be seen that our method successfully recovered the details and also improved the aesthetic quality. In comparison, the other methods failed to produce images with high aesthetic quality, and some even generated unfavorable artifacts that heavily degrade the image. For the quantitative performance comparison, since the SID dataset contains long exposure images, we used them as the ground-truth in measuring the full-reference metrics, \textit{i.e.}, PSNR and SSIM. We also employed several no-reference metrics, \textit{i.e.}, information entropy (IE), global contrast factor (GCF) \cite{matkovic2005global}, and no-reference image quality metric for contrast distortion (NIQMC) \cite{gu2016no}, to evaluate the perceptual image quality. As shown in Table \ref{table_s4c2}, our method shows better performance than the others by a significant margin, indicating that the coefficient estimation branch trained with unpaired data can be well generalized to unseen images. Note that the short-long exposure image pairs in the SID dataset were captured using the same camera configuration except exposure time. Any operation that alters color distribution results in lower PSNR and SSIM.

\subsection{Evaluation of Joint Operation Branch}
The joint enhancement and denoising operation avoids breaking the noise statistics or over-smoothing the input image in comparison with a cascaded application of denoising and enhancement. For the evaluation of the cascaded designs, we tested seven handcrafted or learning-based denoisers, namely BM3D \cite{dabov2007image}, DnCNN \cite{zhang2017beyond}, FFDNet \cite{zhang2018ffdnet}, CBDNet \cite{guo2019toward}, GRDN \cite{kim2019grdn}, SADNet \cite{chang2020spatial}, and CycleISP \cite{zamir2020cycleisp}. For the training of denoisers, we obtained a low-light subset of the SIDD dataset, where we collected samples with the attribute of ``Lowlight'' from SIDD-Small as the test set, and split the rest samples from SIDD-Full into a ratio of 95:5 for training and validation, respectively. All denoising networks were retrained using their default settings.
For the enhancement, we applied the method explained in Section III-C. Specifically, we examined two scenarios: denoising followed by enhancement (dubbed den-enh) and enhancement followed by denoising (dubbed enh-den). The enhancement results without denoising were treated as the baseline. 

\begin{table}[htb]
    \renewcommand{\arraystretch}{1.3}
    \caption{Denoising Results on The SIDD Dataset}
    \label{table_s4d1}
    \centering
    \begin{tabular}{cccccc}
        \hline
        \multirow{2}{*}{Method} & \multirow{2}{*}{Order} & \multicolumn{2}{c}{PSNR} & \multicolumn{2}{c}{SSIM}\\
        & & mean & std & mean & std\\
        \hline
        Baseline & N/A & 24.6866 & 5.2055 & 0.5691 & 0.2308\\
        \multirow{2}{*}{BM3D \cite{dabov2007image}} & den-enh & 33.6075 & 4.9723 & 0.8933 & 0.0950\\
        & enh-den & 24.7832 & 5.2414 & 0.5731 & 0.2311\\
        \multirow{2}{*}{DnCNN \cite{zhang2017beyond}} & den-enh & 32.6793 & 2.9438 & 0.9284 & 0.0715\\
        & enh-den & 30.4027 & \textbf{1.7850} & 0.9373 & 0.0594\\
        \multirow{2}{*}{FFDNet \cite{zhang2018ffdnet}} & den-enh & 33.5070 & 3.9168 & 0.9362 & 0.0616\\
        & enh-den & 33.2565 & 2.2517 & 0.9518 & 0.0495\\
        \multirow{2}{*}{CBDNet \cite{guo2019toward}} & den-enh & 38.0189 & 4.1515 & 0.9574 & 0.0450\\
        & enh-den & 37.5813 & 5.1865 & 0.9502 & 0.0630\\
        \multirow{2}{*}{GRDN \cite{kim2019grdn}} & den-enh & 39.0103 & 4.3996 & 0.9628 & 0.0411\\
        & enh-den & 38.6958 & 4.7934 & 0.9579 & 0.0526\\
        \multirow{2}{*}{SADNet \cite{chang2020spatial}} & den-enh & 38.8018 & 4.3733 & 0.9612 & 0.0434\\
        & enh-den & 38.5222 & 4.8613 & 0.9551 & 0.0578\\
        \multirow{2}{*}{CycleISP \cite{zamir2020cycleisp}} & den-enh & 38.1849 & 4.2101 & 0.9578 & 0.0449\\
        & enh-den & 37.7636 & 5.1127 & 0.9481 & 0.0632\\
        Ours & joint & \textbf{40.4605} & 4.3298 & \textbf{0.9733} & \textbf{0.0315}\\
        \hline
    \end{tabular}
\end{table}

\begin{figure*}[b]
    \centering
    \includegraphics[scale=1]{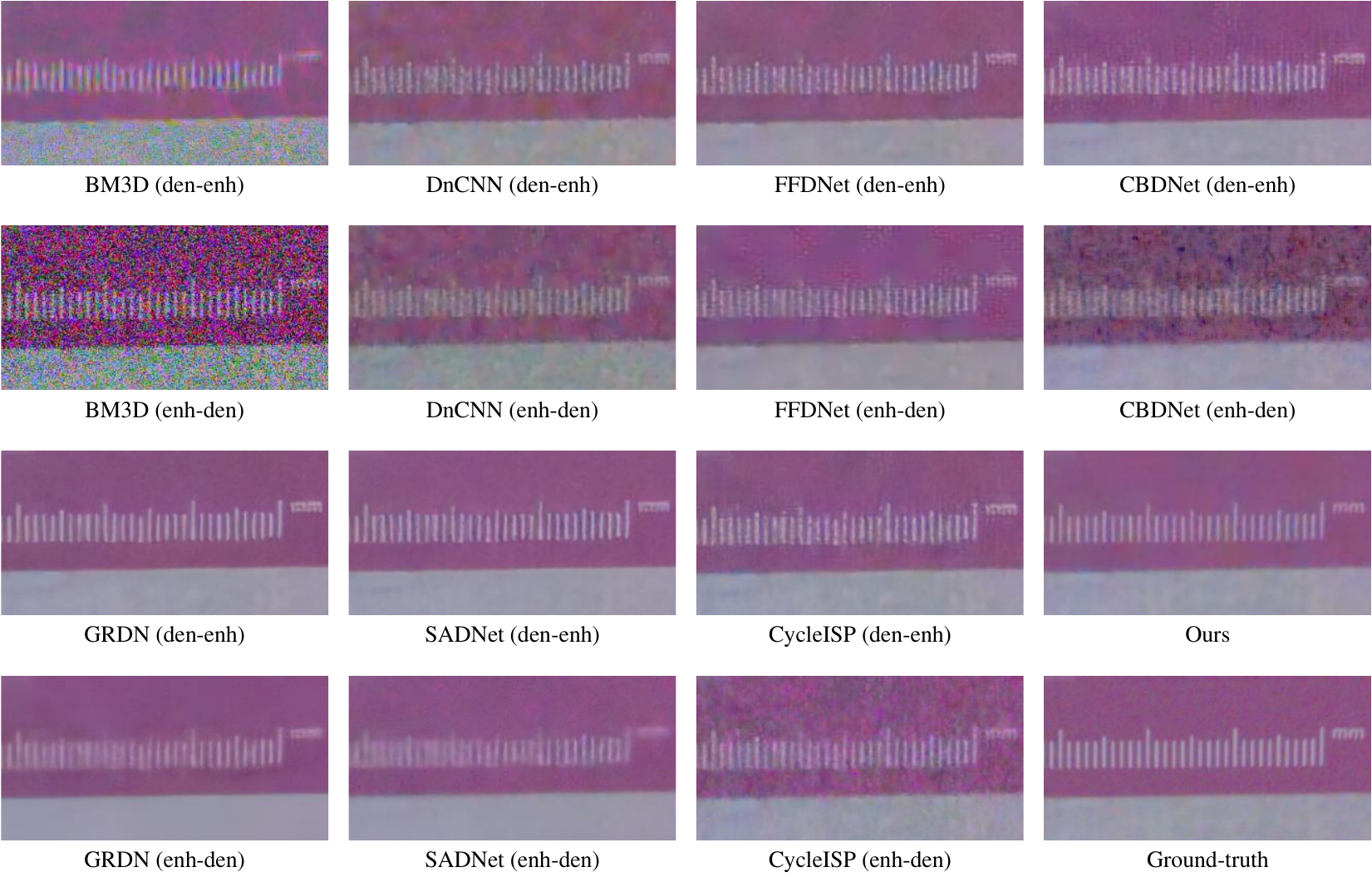}
    \caption{An example of the visual comparison of ours and several state-of-the-art image denoising methods on the SIDD dataset.}
    \label{fig_s4d1}
\end{figure*}

\begin{figure*}[htb]
    \centering
    \includegraphics[scale=1]{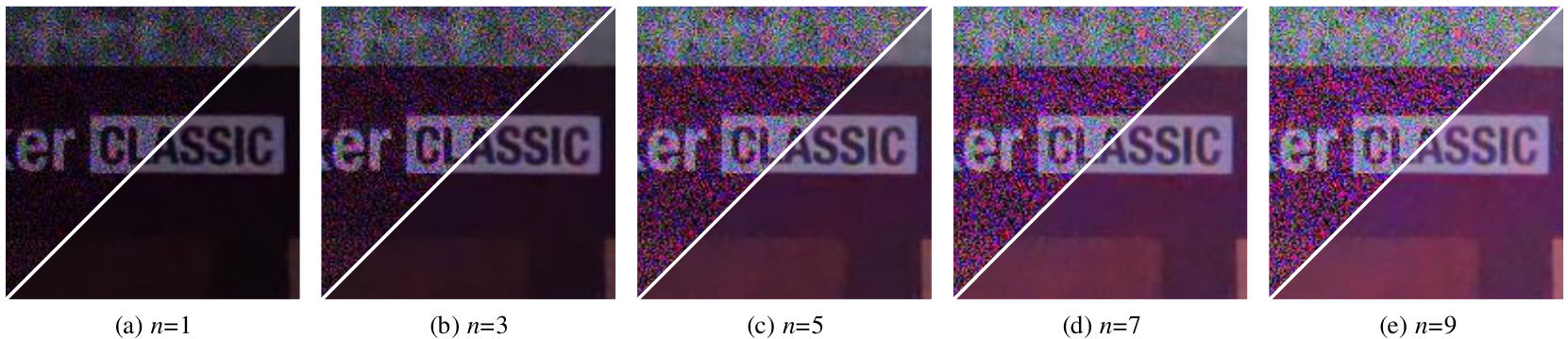}
    \caption{An example of the intermediate enhancement results obtained after $n$ iterations with and without denoising on the SIDD dataset.}
    \label{fig_s4d2}
\end{figure*}

Fig. \ref{fig_s4d1} shows the resultant image examples. It can be seen that cascaded networks either failed in consistent denoising or over-smoothed the image details, while our joint enhancement and denoising better preserved the details and suppressed the noise. Table \ref{table_s4d1} lists the quantitative results, which show evident performance drop in cascaded architectures regardless of processing order. In comparison, our joint method outperforms all the others by a significant margin. Fig. \ref{fig_s4d2} shows the intermediate results, and it is obvious that the noise was heavily amplified during the progressive enhancement without denoising. These results further support our analysis presented in Section III-D.

\subsection{Evaluation of Overall Framework}
To evaluate the complete framework, we made a small dataset using four mobile devices, \textit{i.e.}, DJI Osmo Action, Google Pixel 3, Apple iPhone 11, and Samsung Galaxy S20. Each device was used to capture 20 raw low-light images with various objects, including selfies, buildings, and streets. We used the safe shutter speed to avoid motion blur and limited the ISO below 400 to preserve highlights. In this experiment, the neural image assessment (NIMA) \cite{talebi2018nima}, which is of high correlation to human perception and capable of no-reference image quality assessment in terms of illumination, color, noise level, and high-level aesthetic and semantic features, was used as the evaluation metric. We reported the performance of the proposed framework with eight state-of-the-art learning-based methods, namely Exposure \cite{hu2018exposure}, DeepUPE \cite{wang2019underexposed}, MIRNet \cite{zamir2020learning}, DeepLPF \cite{moran2020deeplpf}, KinD \cite{zhang2019kindling}, RetinexNet \cite{wei2018deep}, EnlightenGAN \cite{jiang2019enlightengan}, and ZeroDCE \cite{guo2020zero}. We estimated $N_{f}$ for each camera model and generated synthetic paired noisy-clean images for training. Each camera model was evaluated individually.

\begin{table*}[b]
    \renewcommand{\arraystretch}{1.3}
    \caption{Joint Enhancement and Denoising Results in Terms of NIMA Scores on Our Dataset}
    \label{table_s4e1}
    \centering
    \begin{tabular}{m{1.25cm}<{\centering}m{1.25cm}<{\centering}m{1.25cm}<{\centering}m{1.25cm}<{\centering}m{1.25cm}<{\centering}m{1.25cm}<{\centering}m{1.25cm}<{\centering}m{1.25cm}<{\centering}m{1.25cm}<{\centering}m{1.25cm}<{\centering}}
    \hline
     Device & Exposure \cite{hu2018exposure} & DeepUPE \cite{wang2019underexposed} & MIRNet \cite{zamir2020learning} & DeepLPF \cite{moran2020deeplpf} & KinD \cite{zhang2019kindling} & RetinexNet \cite{wei2018deep} & Enlgihten \cite{jiang2019enlightengan} & ZeroDCE \cite{guo2020zero} & Ours\\
    \hline
    DJI & 5.6085 & 5.6350 & 5.5286 & 5.7064 & 5.7027 & 4.6945 & 5.8717 & 5.3973 & \textbf{5.9566}\\
    iPhone11 & 4.7139 & 4.8477 & 4.7878 & 4.8585 & 4.7938 & 4.8192 & 4.8157 & 4.7659 & \textbf{5.2594}\\
    Pixel3 & 4.8353 & 4.9247 & 5.0571 & 4.9895 & 4.9639 & 4.6956 & 4.9985 & 4.7611 & \textbf{5.3154}\\
    S20 & 4.8871 & 4.9743 & 5.0065 & 5.0729 & 5.0294 & 4.6520 & 5.1566 & 4.7685 & \textbf{5.3227}\\
    \hline
    mean & 5.0112 & 5.0954 & 5.0950 & 5.1568 & 5.1224 & 4.7153 & 5.2106 & 4.9232 & \textbf{5.4635}\\
    \hline
    \end{tabular}
\end{table*}

\begin{figure*}[b]
    \centering
    \includegraphics[scale=1]{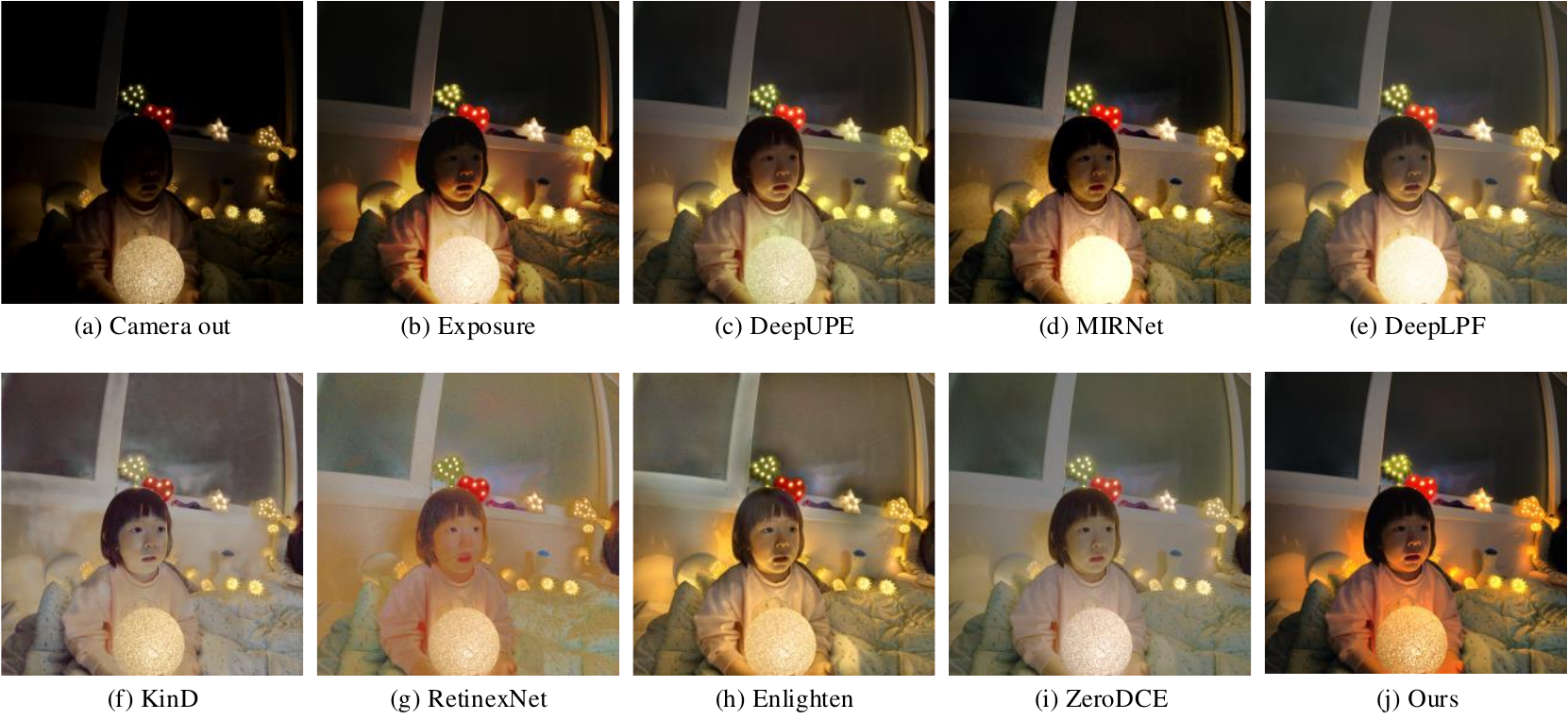}
    \caption{An example of the visual comparison of ours and several state-of-the-art image enhancement methods on our dataset.}
    \label{fig_s4e1}
\end{figure*}

The evaluation results are summarized in Table \ref{table_s4e1}. Our method received the highest NIMA scores among the compared methods and produced consistent performance across four test devices. In contrast, the other methods showed lower performance and were unstable in cross-modal comparison. This can also be observed in Fig. \ref{fig_s4e1}: Even though all the methods improved the image illumination to some extent, most of them did not deliver satisfactory aesthetic quality. For example, Figs. \ref{fig_s4e1}(d) and \ref{fig_s4e1}(e) show over-enhanced bright regions; Figs. \ref{fig_s4e1}(c) and \ref{fig_s4e1}(i) have dull color; Figs. \ref{fig_s4e1}(f) and \ref{fig_s4e1}(h) even contain undesired artifacts. By contrast, our method successfully recovered sufficient details in the dark areas while preserving the highlights and adjusting the overall color and contrast to improve the aesthetic quality.

\begin{figure*}[htb]
    \centering
    \includegraphics[scale=1]{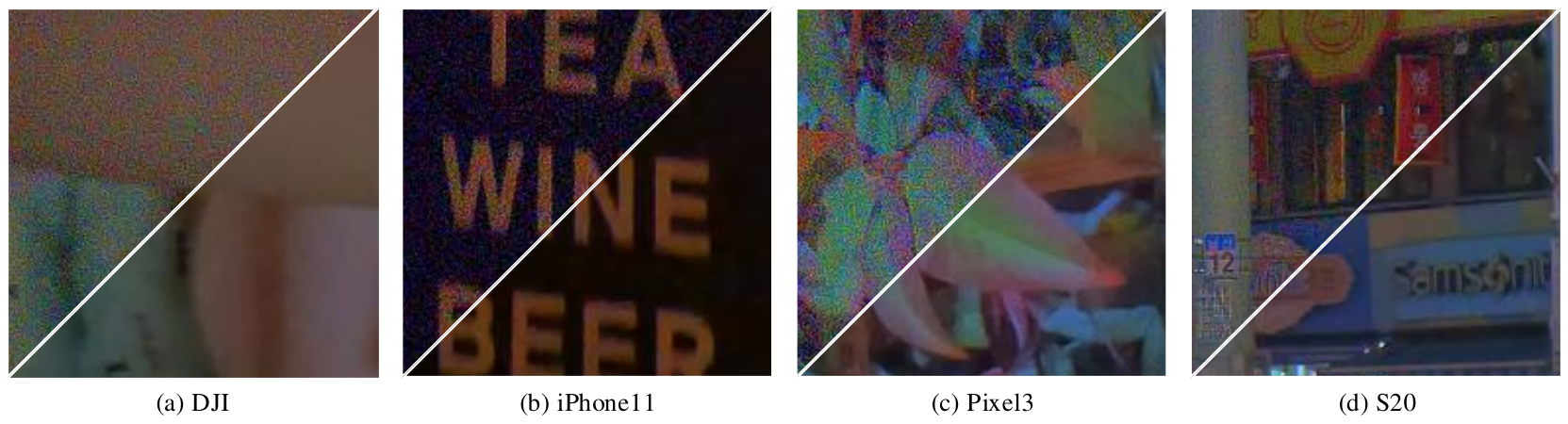}
    \caption{Examples of enhancement results with and without denoising on four camera models.}
    \label{fig_s4e2}
\end{figure*}

Another example of our joint enhancement and denoising is shown in Fig. \ref{fig_s4e2}. The two-branch design significantly relaxed the effort needed in cross-modal noise reduction, resulting in clean images for all the test cameras used. In comparison, the direct enhancement did not consider noise reduction and thus amplified noise, resulting in heavily degraded images that are more challenging for further improvement.

Last, we measured the computation complexity of the proposed method. The average processing time for our dataset (80 images with a resolution of approximately $3,000\times3,000$) is $6.1687$s, which is relatively fast considering our unoptimized code and models. We believe the efficiency can be further improved through network optimization techniques.

\section{Conclusion}
In this paper, we presented a learning-based framework for low-light image enhancement. In contrast to existing works, the proposed framework performs joint illumination enhancement, color enhancement, and model-specific denoising progressively. To make this approach practical in real-world applications, we designed a two-branch structure that handles coefficient estimation, joint enhancement, and denoising in different domains. This design enables it to be retrained for a new camera model using only a few new image samples, significantly reducing the human effort required to recollect massive amounts of paired data. Moreover, the collaborative process allows the network to be more robust and lightweight without compromising performance. Through extensive experiments, we showed that our framework outperforms existing state-of-the-art methods and successfully improves the quality of low-light images. Our future research directions include low-light white-balancing with semantic attention, universal model-independent raw denoising, multi-shot-based low-light imaging, joint low-light image enhancement, denoising, and deblurring.

\section*{Acknowledgment}
The authors would like to thank Dr. Sean Moran for helping reproduce their results and strengthen this paper through constructive discussions. The authors also appreciate the comments received from the reviewers for improving this paper further.

\ifCLASSOPTIONcaptionsoff
  \newpage
\fi

\bibliographystyle{IEEEtran}
\bibliography{IEEEabrv,IEEEexample}

 biography section
\begin{IEEEbiography}[{\includegraphics[width=1in,height=1.25in,clip,keepaspectratio]{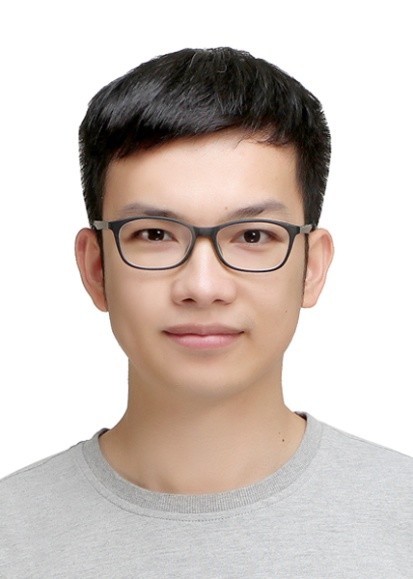}}]{Yucheng Lu} received the B.S. degree in optical information science and technology from Hangzhou Dianzi University, Hangzhou, China, in 2016. He received the Ph.D. degree in multimedia engineering from Department of Multimedia Engineering, Dongguk University, Seoul, Korea, in 2022. He is currently a Research Professor with the BK21 Socialware Information Technology, Korea University. His main research interests include low-level image processing, medical image segmentation and classification, 3D model reconstruction and machine learning based computer vision applications.
\end{IEEEbiography}

\begin{IEEEbiography}[{\includegraphics[width=1in,height=1.25in,clip,keepaspectratio]{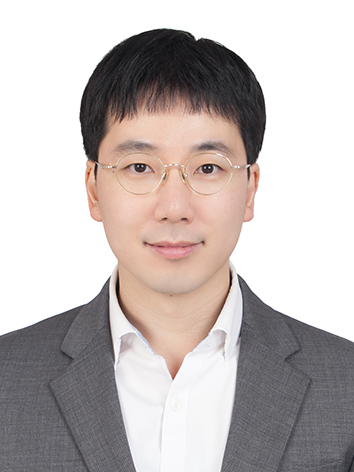}}]{Seung-Won Jung} (S’06-M’11-SM’19) received the B.S. and Ph.D. degrees in electrical engineering from Korea University, Seoul, Korea, in 2005 and 2011, respectively. He was a Research Professor with the Research Institute of Information and Communication Technology, Korea University, from 2011 to 2012. He was a Research Scientist with the Samsung Advanced Institute of Technology, Yongin-si, Korea, from 2012 to 2014. He was an Assistant Professor at the Department of Multimedia Engineering, Dongguk University, Seoul, Korea, from 2014 to 2020. In 2020, he joined the Department of Electrical Engineering at Korea University, where he is currently an Associate Professor. He has published over 70 peer-reviewed articles in international journals. He received the Hae-Dong young scholar award from the Institute of Electronics and Information Engineers in 2019. His current research interests include image processing and computer vision.
\end{IEEEbiography}

\end{document}